\documentclass[aps,showpacs,amsmath,amssymb,12pt]{revtex4}%
\usepackage{amsfonts}
\usepackage{amsmath}
\usepackage{amssymb}
\usepackage{latexsym}
\usepackage{bm}
\usepackage{graphicx}%
\newtheorem{theorem}{Theorem}
\newtheorem{acknowledgement}[theorem]{Acknowledgement}

\begin{document}
\preprint{HEP/123-qed}
\title{Scalar Waves in a Wormhole Topology}
\author{Necmi Bu\u{g}dayc\i}
\email{necmi@dizgeanalitik.com}
\affiliation{Department of Physics, Middle East Technical University, 06531 Ankara, Turkey.}

\begin{abstract}
Global monochromatic solutions of the scalar wave equation are obtained in
flat wormholes of dimensions 2+1 and 3+1. The solutions are in the form of
infinite series involving cylindirical and spherical wave functions and they
are elucidated by the multiple scattering method. Explicit solutions for some
limiting cases are illustrated as well. The results presented in this work
constitute instances of solutions of the scalar wave equation in a spacetime
admitting closed timelike curves.

\end{abstract}
\volumeyear{year}
\volumenumber{number}
\issuenumber{number}
\eid{identifier}
\maketitle

\section{Introduction}

In this paper global solutions of the scalar wave equation in 2+1 and 3+1
dimensional flat wormholes are obtained. As a significant consequence of their
non-trivial topology, wormholes admit closed timelike curves (CTC's). As such
they constitute a suitable framework for the study of the solutions of the
scalar wave equation in a spacetime admitting closed timelike curves. Again
due to the topology of a wormhole, no single coordinate chart is sufficient to
express the global geometry of the whole wormhole spacetime and it becomes
necessary to develop techniques to handle global issues on the one hand and to
investigate the propagation of scalar waves near closed timelike curves. It
should be mentioned that there are works that study the scalar waves that are
valid locally in a certain regions (such as may be termed the
\textquotedblleft throat\textquotedblright) of the wormhole \cite{kar}.

Wormholes are widely studied and discussed, especially after the paper of
Morris and Thorne, in the context of time travel \cite{thorne},\cite{hawk}.
The wave equation attracts attention in this sense that whether the anomalies
of causality violations due to CTC's has corresponding footprints in the
solutions of wave equation.

The metric of the spacetime, as well as the topology, maybe the origin of the
closed timelike curves, as in the case of G\"{o}del's universe \cite{lssu}.
Recently Bachelot has studied the properties of wave equation on a class of
spacetimes of this type \cite{bach}. It may be an interesting question,
whether the origin of the CTC's (being topology induced or metric induced)
effects the global properties of the solutions. Nevertheless, as proposed in
the concluding remarks, the presence of CTC's due to topology does not seem to
have a significant effect on the solutions of wave equation.

Cauchy problem of the scalar wave equation in the flat wormhole considered
here is studied throughout by Friedman and Morris with a variety of other
spacetimes admitting closed timelike curves \cite{FM},\cite{FM1}. They also
proved that there exist a unique solution of Cauchy problem for a class of
spacetimes, including our case, with initial data given at past null infinity
\cite{fr2}.

Due to the wormhole structure, the boundary conditions imposed in solving the
Helmholtz equation depends on the frequency. Therefore spectral theorem is not
applicable in a straight forward manner to express the solution of the wave
equation as a superposition of \ monochromatic wave solutions found in this
work. However, in \cite{fr2}, it is proved using limiting absorption method
that, the superpositions of the monochromatic wave solutions of the problem
converge to the solution of wave equation.

The problem can be handled as a Cauchy problem with given initial data at past
null infinity or alternatively as a scattering problem, i.e. finding scattered
waves from wormhole handle given incident wave.

Our approach is similar to that used in scattering from infinite parallel
cylinders \cite{tw}. $\Psi_{1}$ and $\Psi_{2}$ represents outgoing cylindrical
(or for 3+1 dimensions spherical) waves emerging from the first and the second
wormhole mouth respectively. In order to be able to apply the boundary
conditions conveniently which arise from the peculiar topology of the wormhole
in our case, it is necessary to express $\Psi_{1}$ in terms of cylindrical
(spherical) waves centered at second mouth and vice versa. Addition theorems
for cylindrical and spherical wave functions are employed for this purpose.

The equations for the scattering coefficients of $\Psi_{1}$ and $\Psi_{2}$
that result from the boundary conditions in question are in general not
amenable to direct algebraic manipulation . The multiple scattering method is
applied to obtain an infinite series solution. On the other hand for some
important limiting cases the equations solved explicitly. The solutions by
these both methods are consistent with one other.

The outline of the paper is as follows: In section II, the spacetime is
described and the general formulation of the problem is presented. In section
III, 2+1 dimensional case is studied. The equations are presented, explicit
solutions for two limiting cases are obtained, and finally the multiple
scattering solution is applied. In Section IV the same scheme as section
III\ is followed for 3+1 dimensional case. In section V numerical
verifications of the results obtained in section III\ are presented. Section
VI contains some concluding remarks.

\section{Flat wormhole}

Given a Riemannian manifold $M$, a solution $F:M\times\mathbb{R}%
\rightarrow\mathbb{C}$ of the scalar wave equation
\[
\Delta F=\frac{\partial^{2}F}{\partial t^{2}}%
\]
is said to be a \emph{monochromatic solution} with angular frequency
$\omega\in\mathbb{R}-\{0\}$ if it is of the form $F(m,t)=\Psi(m)e^{i\omega t}$
for some $\Psi:M\rightarrow\mathbb{C}.$ Clearly $\Psi$ is a solution of the
Helmholtz equation
\begin{equation}
\Delta\Psi+\omega^{2}\Psi=0. \label{helm}%
\end{equation}
%

\begin{figure}
[ptb]
\begin{center}
\includegraphics[
width=3.7455in
]%
{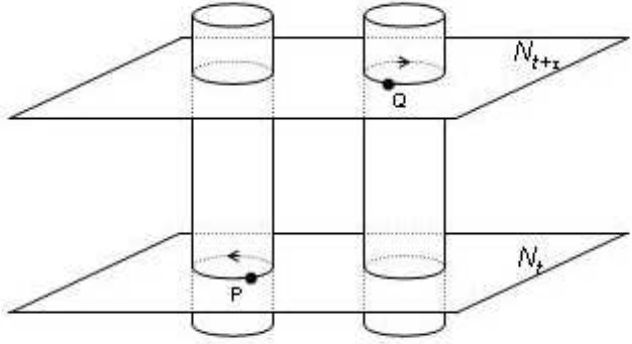}%
\caption{2+1 dimensional flat wormhole. P is identified with Q. Arrows
indicate the direction of the identification.}%
\end{center}
\end{figure}

On a general Lorentzian spacetime the concept of monochromatic solution makes
sense provided the spacetime has an almost product structure that singles out
the time direction locally.

A simple example of a wormhole topology is the flat wormhole described in
\cite{fr2}. This 3+1 dimensional flat wormhole spacetime is constructed as
follows: Let $a,d,\tau\in\mathbb{R}$ with $d>2a>0.$ Consider
\begin{align*}
N  &  =\mathbb{R}^{3}-(\Delta_{+}\cup\Delta_{-}),\\
\hat{M}  &  =N\times\mathbb{R},
\end{align*}
where $\Delta_{+},\Delta_{-}$ are open balls of radius $a>0$ and respective
centers $(0,0,d/2),(0,0,-d/2)$ in $\mathbb{R}^{3}.$ The boundaries of
$\Delta_{+}$ and $\Delta_{-}$ are designated as $\Sigma_{+}$ and $\Sigma_{-}$
respectively. The \emph{wormhole spacetime} $M$ of width $d,$ radius $a,$ and
lag $\tau$ is the semiriemannian manifold obtained as the quotient space of
$\hat{M}$ by identifying events $P,$ $Q$ on $\Sigma_{+}\times\mathbb{R}%
,\Sigma_{-}\times\mathbb{R}$ respectively if $P$ is the reflection of $Q$ in
the $xyt$-plane after a translation by $\tau$ along the $t$-axis, the
semiriemannian metric being naturally inherited from the ordinary Minkowski
metric on $\mathbb{R}^{4}.$ $M$ is clearly a flat Lorentzian spacetime. To be precise:%

\begin{align*}
\Sigma_{+}  &  =\{(x,y,z)\in\mathbb{R}^{3}|x^{2}+y^{2}+(z-d/2)^{2}=a^{2}\},\\
\Sigma_{-}  &  =\{(x,y,z)\in\mathbb{R}^{3}|x^{2}+y^{2}+(z+d/2)^{2}=a^{2}\}.
\end{align*}

For $(x,y,z)\in\Sigma_{+},$ $P$ and $Q$ are identified where%

\begin{align*}
P  &  =(x,y,z,t),\\
Q  &  =(x,y,-z,t+\tau)\text{.}%
\end{align*}

In 2+1 dimensions the manifold is defined in the same way except that:%

\[
N=\left(  ~\mathbb{R}^{2}-(\Delta_{+}\cup\Delta_{-})\right)  ,
\]
$\Delta_{+},\Delta_{-}$ are open disks of radius $a>0$ with respective centers
$(d/2,0),(-d/2,0)$ in $\mathbb{R}^{2}$ and $P$ is the reflection of $Q$ in the
$yt$-plane after a translation by $\tau$ along the $t$-axis.

The geometry for 2+1 dimensions is shown in fig. 1.

Two wormhole conditions arise from this identification map defining the
topology. These conditions will function as boundary conditions imposed on the
general solution of Helmholtz equation in a flat spacetime.

The two wormhole conditions will be denoted as C-1 and C-2. C-1 is%

\[
F(P)=F(Q),
\]
and C-2 is%
\[
\hat{n}_{P}\cdot\nabla F(P)=-\hat{n}_{Q}\cdot\nabla F(Q).
\]
where $\hat{n}_{Q}$ is the unit outward normal to $\Sigma_{-}$ at $Q$ and
$\hat{n}_{P}$ is the unit outward normal to $\Sigma_{+}$ at $P$. In terms of
$\Psi,$ C-1 and C-2 are:%

\[
\Psi(\omega,p)=e^{i\omega\tau}\Psi(\omega,q),
\]

\[
\hat{n}_{P}\cdot\nabla\Psi(\omega,p)=-e^{i\omega\tau}\hat{n}_{Q}\cdot
\nabla\Psi(\omega,q),
\]

\noindent where $p$ and $q$ are the projections of $P$ and $Q$ on $N$ respectively.

The solution will be expressed in three components: An everywhere regular part
of the wave, $\Psi_{0}$, which may be considered as originating from the
sources at past null infinity (or alternatively as the incident wave if the
problem is considered as a scattering problem), and two outgoing waves
originating from each wormhole mouth (or scattered waves from each mouth),
$\Psi_{1}~$\ and $\Psi_{2}$. Obviously $\Psi=\Psi_{0}+\Psi_{1}+\Psi_{2}$

There are two wormhole conditions that enable one to determine two of
$\Psi_{0},\ \Psi_{1}$ and $\Psi_{2}$. The problem will be handled like a
scattering problem and the scattered waves $\Psi_{1}~$\ and $\Psi_{2}$. will
be solved given the incident wave $\Psi_{0}.$

\section{2+1 Dimensions:}

In 2+1 dimensions, solution of Helmholtz equation in cylindrical coordinates
yields Bessel (or Hankel) functions. Being everywhere regular, $\Psi_{0}$ is
expressed in terms of $J_{n}(r),$while $\Psi_{1}$and $\Psi_{2}$ represent
outgoing waves radiating from the wormhole mouths $\Delta_{-}$ and $\Delta
_{+}$, respectively. Outgoing waves are expressed by Hankel functions of the
first kind, $H_{n}^{(1)}(r)$. Referring to fig. 2, $\Psi_{1}$ has its natural
coordinates $(r,\theta)$ centered at $(-d/2,0)$, and $\Psi_{2}$ has its
natural coordinates $(R,\phi)$ centered at $(d/2,0)$. The coordinate
variables, $\theta$ and $\phi$ are chosen in this way to make use of the
mirror symmetry of the geometry of the wormhole with respect to $y$ axis.
Since $\Psi_{0},$ $\Psi_{1}$and $\Psi_{2}$ are valid in exterior domain, they
are expressed in terms of integer order Bessel (Hankel) functions only.
Therefore the expansion of $\Psi_{0}$, $\Psi_{1}$and $\Psi_{2}$ in terms of
Bessel (Hankel) functions are:%

\begin{figure}
[ptb]
\begin{center}
\includegraphics[
width=3.8614in
]%
{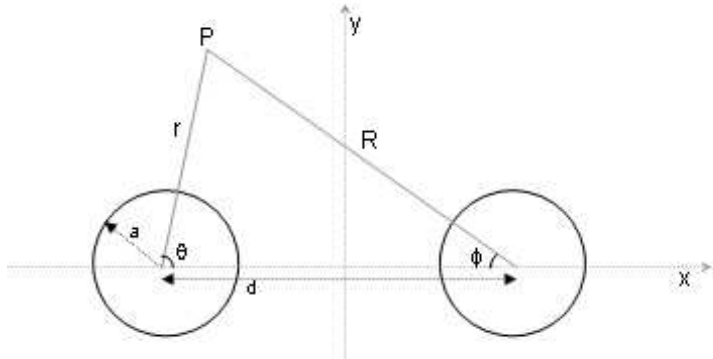}%
\caption{Coordinates used for 2+1 dimensions.}%
\end{center}
\end{figure}
%

\begin{align*}
\Psi_{0}  &  =\sum\limits_{n=-\infty}^{\infty}A_{n}J_{n}(\omega r)e^{in\theta
},\\
\Psi_{1}  &  =\sum\limits_{n=-\infty}^{\infty}B_{n}H_{n}^{(1)}(\omega r)\cdot
e^{in\theta},\\
\Psi_{2}  &  =\sum\limits_{n=-\infty}^{\infty}C_{n}H_{n}^{(1)}(\omega R)\cdot
e^{in\phi}.
\end{align*}

$B_{n}$ and$\ C_{n}$ will be found given the coefficients of the incident wave
$A_{n}$. The two wormhole conditions supply the two equations to determine the
unknown coefficients $B_{n}$ and $\ C_{n}.$

The wormhole conditions C-1 and C-2 are:%

\begin{align*}
\Psi|_{R=a,\phi=\theta}  &  =e^{i\omega\tau}\Psi|_{r=a,\theta}\ \ \ \ \ -\pi
<\theta\leq\pi,\\
\dfrac{\partial}{\partial R}\Psi|_{R=a,\phi=\theta}  &  =-e^{i\omega\tau
}\dfrac{\partial}{\partial r}\Psi|_{r=a,\theta}\ \ \ \ \ -\pi<\theta\leq\pi.
\end{align*}
$\qquad$

To compute $\Psi$ at $R=a$ and $r=a$ it is necessary to write down $\Psi_{0},$
$\Psi_{1}$ in $(R,\phi)$ coordinates and $\Psi_{2}$ in $(r,\theta)$
coordinates. The addition theorem for cylindrical harmonics is used for
expressing a cylindrical wave in terms of cylindrical waves of a translated
origin \cite{watson}. It should be noted that, unlike the everywhere regular
Bessel functions $J_{n}(\omega r),$ there are two different versions of the
addition theorems of Hankel functions. For $\vec{r}=\vec{d}+\vec{R},$ addition
theorems yield%

\begin{align}
H_{n}^{(1)}(\omega R)e^{in(\pi-\phi)}  &  =\left\{
\begin{array}
[c]{c}%
\sum\limits_{k=-\infty}^{\infty}J_{k}(\omega d)H_{n+k}^{(1)}(\omega
r)e^{i(n+k)\theta}\ \ \ \text{if}\ \ r>d\\
\sum\limits_{k=-\infty}^{\infty}H_{k}^{(1)}(\omega d)J_{n+k}(\omega
r)e^{i(n+k)\theta}\ \ \text{\ if\ }\ r<d
\end{array}
\right.  ,\label{hd}\\
H_{n}^{(1)}(\omega r)e^{in(\theta-\pi)}  &  =\left\{
\begin{array}
[c]{c}%
\sum\limits_{k=-\infty}^{\infty}J_{k}(\omega d)H_{n+k}^{(1)}(\omega
R)e^{-i(n+k)\phi}\ \ \ \text{if}\ \ R>d\\
\sum\limits_{k=-\infty}^{\infty}H_{k}^{(1)}(\omega d)J_{n+k}(\omega
R)e^{-i(n+k)\phi}\ \ \text{\ if\ }\ R<d
\end{array}
\right.  ,\label{hd2}\\
J_{n}(\omega r)e^{in(\theta-\pi)}  &  =\sum\limits_{k=-\infty}^{\infty}%
J_{k}(\omega d)J_{n+k}(\omega R)e^{i(n+k)\phi}. \label{jadd}%
\end{align}

Wormhole conditions require the expression at $r=a$ and $R=a$. Since $a<d,$
$r<d$ versions of (\ref{hd}) and (\ref{hd2}) should be used.

Accordingly, the wave functions are expressed as a sum of Bessel functions at
translated origin as%
\begin{align*}
\sum\limits_{n=-\infty}^{\infty}A_{n}J_{n}(\omega r)e^{in\theta}  &
=\sum\limits_{n=-\infty}^{\infty}\bar{A}_{n}\cdot J_{n}(\omega R)e^{in\phi},\\
\sum\limits_{n=-\infty}^{\infty}B_{n}H_{n}^{(1)}(\omega r)e^{in\theta}  &
=\sum\limits_{n=-\infty}^{\infty}\bar{B}_{n}\cdot J_{n}(\omega R)\cdot
e^{in\phi},\\
\sum\limits_{n=-\infty}^{\infty}C_{n}H_{n}^{(1)}(\omega R)e^{in\phi}  &
=\sum\limits_{n=-\infty}^{\infty}\bar{C}_{n}\cdot J_{n}(\omega r)\cdot
e^{in\theta}.\\
&
\end{align*}

The expressions for $\bar{A}_{n}$, $\bar{B}_{n}$ and $\bar{C}_{n}$ are found
using (\ref{hd}), (\ref{hd2}) and (\ref{jadd}) as:%

\begin{align}
\bar{A}_{n}  &  =\sum\limits_{k=-\infty}^{\infty}A_{k-n}J_{k}(\omega
d),\label{t2a}\\
\bar{B}_{n}  &  =\sum\limits_{k=-\infty}^{\infty}B_{k-n}H_{k}^{(1)}(\omega
d),\label{t2b}\\
\bar{C}_{n}  &  =\sum\limits_{k=-\infty}^{\infty}C_{k-n}H_{k}^{(1)}(\omega d).
\label{t2c}%
\end{align}

Having obtained the expression of the wave in the coordinates centered at each
mouth, application of wormhole conditions give necessary equations for the
unknown coefficients $B_{n}$ and $C_{n}$.

C-1 leads to%
\begin{multline*}
\sum\limits_{n=-\infty}^{\infty}(A_{n}\cdot J_{n}(\omega a)+B_{n}\cdot
H_{n}^{(1)}(\omega a)+\bar{C}_{n}J_{n}(\omega a))e^{in\theta}\\
=e^{i\omega\tau}\sum\limits_{n=-\infty}^{\infty}(\bar{A}_{n}J_{n}(\omega
a)+\bar{B}_{n}J_{n}(\omega a)+C\cdot H_{n}^{(1)}(\omega a))e^{in\theta},
\end{multline*}

\begin{equation}
B_{n}-e^{i\omega\tau}C_{n}=-\dfrac{J_{n}(\omega a)}{H_{n}^{(1)}(\omega
a)}(A_{n}-e^{i\omega\tau}\bar{A}_{n}+\bar{C}_{n}-e^{i\omega\tau}\bar{B}_{n}),
\label{c12}%
\end{equation}
and C-2 leads to%

\begin{multline*}
\sum\limits_{n=-\infty}^{\infty}\left(  A_{n}\cdot\dfrac{\partial}{\partial
r}J_{n}(\omega r)|_{r=a}+B_{n}\cdot\dfrac{\partial}{\partial r}H_{n}%
^{(1)}(\omega r)|_{r=a}+\bar{C}_{n}\dfrac{\partial}{\partial r}J_{n}(\omega
r)|_{r=a}\right)  e^{in\theta}\\
=-e^{i\omega\tau}\sum\limits_{n=-\infty}^{\infty}\left(  \bar{A}_{n}%
\dfrac{\partial}{\partial r}J_{n}(\omega r)|_{r=a}+\bar{B}_{n}\dfrac{\partial
}{\partial r}J_{n}(\omega r)|_{r=a}+C_{n}\dfrac{\partial}{\partial r}%
H_{n}^{(1)}(\omega r)|_{r=a}\right)  e^{in\theta},
\end{multline*}

\begin{equation}
B_{n}+e^{i\omega\tau}C_{n}=-\dfrac{\dfrac{\partial}{\partial r}J_{n}(\omega
r)|_{r=a}}{\dfrac{\partial}{\partial r}H_{n}^{(1)}(\omega r)|_{r=a}}%
(A_{n}+e^{i\omega\tau}\bar{A}_{n}+\bar{C}_{n}+e^{i\omega\tau}\bar{B}_{n}).
\label{c22}%
\end{equation}
$\qquad$\bigskip

Solving (\ref{c12}) and (\ref{c22}) for $B_{n}$ and $C_{n}$, one finds%
\begin{align}
B_{n}  &  =-\gamma_{n}^{+}(\omega a)\bar{C}_{n}+e^{i\omega\tau}\gamma_{n}%
^{-}(\omega a)\bar{B}_{n}-\gamma_{n}^{+}(\omega a)A_{n}+e^{i\omega\tau}%
\gamma_{n}^{-}(\omega a)\bar{A}_{n},\label{2da}\\
C_{n}  &  =-\gamma_{n}^{+}(\omega a)\bar{B}_{n}+e^{-i\omega\tau}\gamma_{n}%
^{-}(\omega a)\bar{C}_{n}-\gamma_{n}^{+}(\omega a)\bar{A}_{n}+e^{-i\omega\tau
}\gamma_{n}^{-}(\omega a)A_{n}, \label{2db}%
\end{align}
where%

\begin{align*}
\gamma_{n}^{+}(\omega a)  &  \triangleq\dfrac{1}{2}\left(  \dfrac{J_{n}(\omega
a)}{H_{n}^{(1)}(\omega a)}+\dfrac{\dfrac{\partial}{\partial r}J_{n}(\omega
r)|_{r=a}}{\dfrac{\partial}{\partial r}H_{n}^{(1)}(\omega r)|_{r=a}}\right)
,\\
\gamma_{n}^{-}(\omega a)  &  \triangleq\dfrac{1}{2}\left(  \dfrac{J_{n}(\omega
a)}{H_{n}^{(1)}(\omega a)}-\dfrac{\dfrac{\partial}{\partial r}J_{n}(\omega
r)|_{r=a}}{\dfrac{\partial}{\partial r}H_{n}^{(1)}(\omega r)|_{r=a}}\right)  .
\end{align*}

For the sake of simplicity the known parts of (\ref{2da}) and (\ref{2db}) will
be denoted by $E_{n}$ and $F_{n}$ respectively.%

\begin{align}
E_{n}  &  =-\gamma_{n}^{+}(\omega a)A_{n}+e^{i\omega\tau}\gamma_{n}^{-}(\omega
a)\bar{A}_{n},\label{en}\\
F_{n}  &  =-\gamma_{n}^{+}(\omega a)\bar{A}_{n}+e^{-i\omega\tau}\gamma_{n}%
^{-}(\omega a)A_{n}. \label{fn}%
\end{align}

This pair of equations (\ref{2da}) and (\ref{2db}) are not solvable
explicitly; however it is possible to solve them for the limiting cases $a\ll
d\ $and $a\ll1$.

\subsection{Solutions for $a\ll d\ $and $a\ll1:$}

The difficulty in solving (\ref{2da}) and (\ref{2db}) arises from
the\ convolution sum present in the expressions of $\bar{B}_{n}$ and $\bar
{C}_{n}.$ However this term can be evaluated for special forms of $H_{n}%
^{(1)}(\omega d)$, namely when it is in complex exponential $e^{-in\alpha}$
form. When $|n|\ll\omega d$, asymptotically $\ H_{n}^{(1)}(\omega d)$ becomes
$e^{-in\pi/2}$ as a function of $n$. $\gamma_{n}^{+}(\omega a)$ and
$\gamma_{n}^{-}(\omega a)$ are almost zero for $|n|\gtrsim2\omega a,$ and so
are $B_{n}$ and $\ C_{n}.$ Thus when $a\ll d,$ the only terms that contribute
to $\gamma_{n}^{\pm}(\omega a)\bar{B}_{n}$ ($\gamma_{n}^{\pm}(\omega a)\bar
{C}_{n}$) are those satisfy $|n|\lesssim2\omega a\ll\omega d$. The $a\ll d$
case is of practical importance in physics. In a wormhole universe, this
corresponds to the case that the wormhole is connecting regions of the
universe that are spatially far from each other compared to the radius of the wormhole.

This approximation is not valid for the high frequency limit in general.

When $a\ll1,$ $\gamma_{n}^{\pm}(\omega a)$ tends to zero unless $n\neq0$,
regardless of $d$. Accordingly, so are $B_{n}$ and $\ C_{n}.$

These two cases in which approximate solutions\ are possible, $a\ll d$ and
$a\ll1$, are examined below.

\subsubsection{$a\ll d$}

For large $\omega d,$ asymptotic formula for $H_{n}^{(1)}(\omega d)$ is%

\begin{align}
H_{n}^{(1)}(\omega d)  &  =z(\omega d)e^{-in\pi/2}\label{happ}\\
&  \times(1+i\dfrac{4n^{2}-1}{1!(8\omega d)}+i^{2}\dfrac{(4n^{2}-1)(4n^{2}%
-9)}{2!(8\omega d)^{2}}+i^{3}\dfrac{(4n^{2}-1)(4n^{2}-9)(4n^{2}-25)}%
{3!(8\omega d)^{3}}+...),\nonumber
\end{align}

\[
z(\omega d)\triangleq\sqrt{\dfrac{2}{\pi\omega d}}e^{i(\omega d-(\pi/4))}.
\]

For $n^{2}\ll\omega d,\ $\ the infinite sum inside the brackets can be
approximated to $1:$%

\[
H_{n}^{(1)}(\omega d)\approx z(\omega d)e^{-in\pi/2}%
\]

This form of $H_{n}^{(1)}(\omega d)$ allows one to evaluate the sum $\bar
{B}_{n}$:%

\begin{align*}
\sum\limits_{k=-\infty}^{\infty}B_{(k-n)}H_{k}^{(1)}(\omega d)  &
=\sum\limits_{m=-\infty}^{\infty}B_{m}H_{n+m}^{(1)}(\omega d)\approx z(\omega
d)(\sum\limits_{m=-\infty}^{\infty}B_{m}e^{-im\pi/2})e^{-in\pi/2}\\
&  =z(\omega d)\hat{B}(\pi/2)e^{-in\pi/2}%
\end{align*}
where hat denotes the Fourier sum:%

\[
\hat{X}(\alpha)\triangleq\sum\limits_{m=-\infty}^{\infty}X_{m}e^{-im\alpha}.
\]

Substituting into (\ref{2da}) and (\ref{2db})%

\begin{align}
B_{n}  &  =z(\omega d)[-\hat{C}(\pi/2)\gamma_{n}^{+}+e^{i\omega\tau}\hat
{B}(\pi/2)\gamma_{n}^{-}]e^{-in\pi/2}+E_{n},\label{agd_b}\\
C_{n}  &  =z(\omega d)[-\hat{B}(\pi/2)\gamma_{n}^{+}+e^{-i\omega\tau}\hat
{C}(\pi/2)\gamma_{n}^{-}]e^{-in\pi/2}+F_{n}. \label{agd_c}%
\end{align}

The right hand sides of (\ref{agd_b}) and (\ref{agd_c}) involves $\hat{C}%
(\pi/2)$ and $\tilde{B}(\pi/2)$ which are unknown yet. Multiplying each side
by $e^{-in\pi/2}$ and sum over $n$'s gives a pair of equations for $\hat
{C}(\pi/2)$ and $\tilde{B}(\pi/2)$ :%

\begin{equation}%
\begin{bmatrix}
\hat{B}(\pi/2)\\
\hat{C}(\pi/2)
\end{bmatrix}
=%
\begin{bmatrix}
1-z(\omega d)e^{i\omega\tau}\tilde{\gamma}^{-}(\pi) & z(\omega d)\tilde
{\gamma}^{+}(\pi)\\
z(\omega d)\tilde{\gamma}^{+}(\pi) & 1-z(\omega d)e^{-i\omega\tau}%
\tilde{\gamma}^{-}(\pi)
\end{bmatrix}
^{-1}%
\begin{bmatrix}
\hat{E}(\pi/2)\\
\hat{F}(\pi/2)
\end{bmatrix}
\label{ad}%
\end{equation}

The numerical results comparing the solutions obtained by these formulae and
by the multiple scattering method is presented in the appendix.

To have a better approximation, the second term $i\dfrac{4n^{2}-1}{1!(8\omega
d)}\ $in the infinite sum of in (\ref{happ})\ can be included:%

\[
H_{n}^{(1)}(\omega d)\ \ \approx\sqrt{\dfrac{2}{\pi\omega d}}e^{i(\omega
d-(\pi/4))}e^{-in\pi/2}(1+i\dfrac{4n^{2}-1}{1!(8z)})
\]

In this case, the expression for $H_{n}^{(1)}(\omega d)$ involves
$n^{2}e^{-in\pi/2}.$ This form of $H_{n}^{(1)}(\omega d)$ still enables one to
evaluate $\bar{B}_{n}$ and $\bar{C}_{n},$ explicitly. This time there arise
$6$ unknowns and (\ref{ad}) is replaced by a $6$x$6$ matrix equation. In this
way it is possible to have better approximations by taking more terms into
account in (\ref{happ}). The number of the linear equations is $4k-2$ when the
first $k$ term is taken into account in (\ref{happ}).

\subsubsection{${\protect\large a\ll1:}$}

At $\omega a=0,$ the Bessel function $J_{n}(\omega a)$ is a discrete delta
function with respect to variable $n,$ and its derivative is zero for all $n:$%

\begin{equation}
J_{n}(0)=\left\{
\begin{array}
[c]{l}%
1\text{ if }n=0\\
0\text{ otherwise}%
\end{array}
\right.  ;\text{ \ \ \ }\dfrac{\partial}{\partial r}J_{n}(\omega
r)|_{r=0}=0\text{ for all }n.
\end{equation}

Therefore, in the limit $a$ goes to zero, $\gamma_{n}^{+}(\omega a)$ and
$\gamma_{n}^{-}(\omega a)$ become discrete delta functions:%

\[
\gamma_{n}^{+}(\omega a)\approx\gamma_{n}^{-}(\omega a)\approx\dfrac
{J_{0}(\omega a)}{H_{0}^{(1)}(\omega a)}\delta_{n},
\]
where \cite{jackson},%

\begin{equation}
\dfrac{J_{0}(\omega a)}{H_{0}^{(1)}(\omega a)}\triangleq\gamma_{0}=\dfrac
{1}{1+\dfrac{2i}{\pi}(\ln(\dfrac{\omega a}{2})+0.5772)}.
\end{equation}

The $\gamma_{n}^{\pm}(\omega a)$ factors standing in front of each term on the
right hand sides of (\ref{2da}) and (\ref{2db}) make $B_{n}$ and $C_{n}$ delta
functions as well.
\begin{align*}
B_{n}  &  =B_{0}\delta_{n},\\
C_{n}  &  =C_{0}\delta_{n};
\end{align*}
so that,%

\begin{align*}
\Psi_{1}  &  =B_{0}\cdot H_{0}^{(1)}(\omega r),\\
\Psi_{2}  &  =C_{0}\cdot H_{0}^{(1)}(\omega R).
\end{align*}

$B_{0}$ and $C_{0}$ are found by substitution to (\ref{2da}) and (\ref{2db}):%

\begin{align*}
\gamma_{n}^{\pm}(\omega a)\bar{B}_{n}  &  \approx(\gamma_{0}\delta_{n}%
)(B_{0}H_{n}^{(1)}(\omega d))=\gamma_{0}B_{0}H_{0}^{(1)}(\omega d)\delta
_{n},\\
\gamma_{n}^{\pm}(\omega a)\bar{C}_{n}  &  \approx\gamma_{0}C_{0}H_{0}(\omega
d)\delta_{n},
\end{align*}

\[%
\begin{bmatrix}
B_{0}\\
C_{0}%
\end{bmatrix}
=%
\begin{bmatrix}
1-\gamma_{0}H_{0}(\omega d)e^{i\omega\tau} & \gamma_{0}H_{0}(\omega d)\\
\gamma_{0}H_{0}(\omega d) & 1-\gamma_{0}H_{0}(\omega d)e^{-i\omega\tau}%
\end{bmatrix}
^{-1}%
\begin{bmatrix}
E_{0}\\
F_{0}%
\end{bmatrix}
,
\]
where%

\begin{align*}
E_{0}  &  =\gamma_{0}(-A_{0}+\ e^{i\omega\tau}\sum\limits_{m=-\infty}^{\infty
}A_{m}J_{m}(\omega d)),\\
F_{0}  &  =\gamma_{0}(-\sum\limits_{m=-\infty}^{\infty}A_{m}J_{m}(\omega
d)+\ e^{-i\omega\tau}A_{0}).
\end{align*}

\subsection{Multiple scattering}

An alternative approach is the use of multiple scattering \cite{tw}%
,\cite{heavi}. In the multiple scattering method, the scattered waves are
decomposed into lower order scattered waves from each mouth. Initially each
wormhole mouth is considered to be excited by only the incident wave and first
order scattering coefficients are found by imposing wormhole conditions. Then
each wormhole is considered to be excited by only the first order scattered
wave from the other mouth and the second order scattering coefficient are
found imposing wormhole conditions. $k^{th}$ order scattering coefficients are
found by continuing the same procedure. The scattered wave from each mouth is
the sum of these $k^{th}$ order scattering coefficients.%

\[
\Psi_{1}=\sum\limits_{k=1}^{\infty}\Psi_{1}^{k}%
\]
where%

\[
\Psi_{1}^{k}=\sum\limits_{n=-\infty}^{\infty}B_{n}^{k}H_{n}^{(1)}(\omega
r)\cdot e^{in\theta}\text{ and }B_{n}=\sum\limits_{k=1}^{\infty}B_{n}^{k}.
\]

Similarly,%

\begin{align*}
\Psi_{2}  &  =\sum\limits_{k=1}^{\infty}\Psi_{2}^{k},\\
\Psi_{2}^{k}  &  =\sum\limits_{n=-\infty}^{\infty}C_{n}^{k}H_{n}^{(1)}(\omega
R)\cdot e^{in\phi}.
\end{align*}

Wormhole conditions for the first order scattering coefficients are:%

\begin{align}
(\Psi_{0}+\Psi_{1}^{1})_{r=a}  &  =e^{i\omega\tau}(\Psi_{0}+\Psi_{2}%
^{1})_{R=a,\phi=\theta},\label{ms2a}\\
\dfrac{\partial}{\partial r}(\Psi_{0}+\Psi_{1}^{1})|_{r=a}  &  =-e^{i\omega
\tau}\dfrac{\partial}{\partial R}(\Psi_{0}+\Psi_{2}^{1})|_{R=a,\phi=\theta}.
\label{ms2b}%
\end{align}

Similarly for the $(k+1)^{th}$ order coefficients:%

\begin{align*}
(\Psi_{1}^{k+1}+\Psi_{2}^{k})_{r=a}  &  =e^{i\omega\tau}(\Psi_{1}^{k}+\Psi
_{2}^{k+1})|_{R=a,\phi=\theta},\\
\dfrac{\partial}{\partial r}(\Psi_{1}^{k+1}+\Psi_{2}^{k})|_{r=a}  &
=-e^{i\omega\tau}\dfrac{\partial}{\partial R}(\Psi_{1}^{k}+\Psi_{2}%
^{k+1})|_{R=a,\phi=\theta}.
\end{align*}

It is easy to see that when $1^{st}$ and $k^{th}$ order scattered waves
satisfy wormhole conditions, total scattered wave satisfies as well.%

\begin{multline*}
\Psi|_{r=a}=(\Psi_{0}+\Psi_{1}+\Psi_{2})|_{r=a}=(\Psi_{0}+\Psi_{1}^{1}%
+\sum\limits_{k=1}^{\infty}(\Psi_{1}^{k+1}+\Psi_{2}^{k}))|_{r=a}\\
=e^{i\omega\tau}(\Psi_{0}+\Psi_{2}^{1})|_{R=a,\phi=\theta}+e^{i\omega\tau}%
\sum\limits_{k=1}^{\infty}(\Psi_{l}^{k}+\Psi_{2}^{k+1})|_{R=a,\phi=\theta
}=e^{i\omega\tau}(\Psi_{0}+\Psi_{1}+\Psi_{2})|_{R=a,\phi=\theta}.
\end{multline*}

Imposing wormhole conditions to (\ref{ms2a}) and (\ref{ms2b}), first order
scattering coefficients are obtained.

C-1 yields:%

\[
\sum\limits_{n=-\infty}^{\infty}A_{n}\cdot J_{n}(\omega a)e^{in\theta}%
+B_{n}^{1}\cdot H_{n}^{(1)}(\omega a)\cdot e^{in\theta}=e^{i\omega\tau}%
(\sum\limits_{n=-\infty}^{\infty}\bar{A}_{n}J_{n}(\omega a)e^{in\theta}%
+\sum\limits_{n=-\infty}^{\infty}C_{n}^{1}H_{n}^{(1)}(\omega a)\cdot
e^{in\theta}),
\]

\[
B_{n}^{1}-e^{i\omega\tau}C_{n}^{1}=-\dfrac{J_{n}(\omega a)}{H_{n}^{(1)}(\omega
a)}(A_{n}-e^{i\omega\tau}\bar{A}_{n}).
\]

C-2 yields:%

\[
B_{n}^{1}+e^{i\omega\tau}C_{n}^{1}=-\dfrac{\dfrac{\partial}{\partial r}%
J_{n}(\omega r)|_{r=a}}{\dfrac{\partial}{\partial r}H_{n}^{(1)}(\omega
r)|_{r=a}}(A_{n}+e^{i\omega\tau}\bar{A}_{n}).
\]
\bigskip

Solving for $B_{n}^{1}$ and $C_{n}^{1}:$%
\begin{align}
B_{n}^{1}  &  =-\gamma_{n}^{+}(\omega a)A_{n}+e^{i\omega\tau}\gamma_{n}%
^{-}(\omega a)\bar{A}_{n},\label{ms2a1}\\
C_{n}^{1}  &  =-\gamma_{n}^{+}(\omega a)\bar{A}_{n}+e^{-i\omega\tau}\gamma
_{n}^{-}(\omega a)A_{n}. \label{ms2b1}%
\end{align}

Note that $B_{n}^{1}$ and $C_{n}^{1}$ are equal to the known parts of
(\ref{2da}) and (\ref{2db}), $E_{n}$ and $F_{n},$ respectively.

$k^{th}$order scattering coefficients are obtained similarly as:%

\begin{align}
B_{n}^{k+1}  &  =-\gamma_{n}^{+}(\omega a)\bar{C}_{n}^{k}+e^{i\omega\tau
}\gamma_{n}^{-}(\omega a)\bar{B}_{n}^{k},\label{ms2ak}\\
C_{n}^{k+1}  &  =-\gamma_{n}^{+}(\omega a)\bar{B}_{n}^{k}+e^{-i\omega\tau
}\gamma_{n}^{-}(\omega a)\bar{C}_{n}^{k}. \label{ms2bk}%
\end{align}
%

\begin{figure}
[ptb]
\begin{center}
\includegraphics[
width=1.7443in
]%
{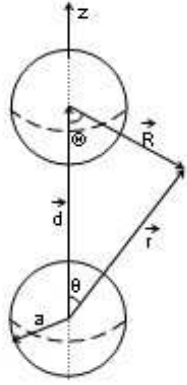}%
\caption{Coordinates used for 3+1 dimensions.}%
\end{center}
\end{figure}

\section{\bigskip3+1 DIMENSIONS}

In 3+1 dimensions, the solutions of wave equation in spherical coordinates,
i.e. spherical wave functions, involve spherical Bessel functions and
spherical harmonics \cite{strat}. In agreement with the 2+1 dimensional case,
$\Psi_{0}$ is expressed in terms of usual spherical Bessel functions, while
$\Psi_{1}$\ and $\Psi_{2}$ are expressed in terms of spherical Hankel
functions. Referring to fig.3,%

\begin{align}
\Psi_{0}  &  =\sum\limits_{l=-\infty}^{\infty}\sum\limits_{m=-l}^{l}%
A_{lm}\cdot j_{l}(\omega r)Y_{lm}(\theta,\varphi),\label{3jadd}\\
\Psi_{1}  &  =\sum\limits_{l=-\infty}^{\infty}\sum\limits_{k=-l}^{l}%
B_{lm}\cdot h_{l}^{(1)}(\omega r)Y_{lm}(\theta,\varphi),\label{3hr}\\
\Psi_{2}  &  =\sum\limits_{l=-\infty}^{\infty}\sum\limits_{m=-l}^{l}%
C_{lm}\cdot h_{l}^{(1)}(\omega R)Y_{lm}(\Theta,\varphi), \label{3hR}%
\end{align}
and the wormhole conditions are,%

\begin{align*}
\Psi|_{r=a,\theta,\varphi}  &  =e^{i\omega\tau}\Psi|_{R=a,\Theta
=\theta,\varphi};\qquad0\leq\theta\leq\pi,\ -\pi<\varphi\leq\pi,\\
\dfrac{\partial}{\partial r}\Psi|_{r=a,\theta,\varphi}  &  =-e^{i\omega\tau
}\dfrac{\partial}{\partial R}\Psi|_{R=a,\Theta=\theta,\varphi;}\qquad
0\leq\theta\leq\pi,\ -\pi<\varphi\leq\pi.
\end{align*}

The addition theorems for the spherical wave functions, for $\vec{r}=\vec
{d}+\vec{R}$ are \cite{feld},\cite{russek}:%

\begin{align}
j_{l}(\omega r)Y_{lm}(\theta,\varphi)  &  =\sum\limits_{l^{\prime}m^{\prime}%
}\alpha_{l^{\prime}m^{\prime}}^{lm+}(\vec{d})j_{l^{\prime}}(\omega
R)Y_{l^{\prime}m^{\prime}}(\pi-\Theta,\varphi),\label{ad3a}\\
h_{l}^{(1)}(\omega r)Y_{lm}(\theta,\varphi)  &  =\sum\limits_{l^{\prime
}m^{\prime}}\alpha_{l^{\prime}m^{\prime}}^{lm}(\vec{d})j_{l^{\prime}}(\omega
R)Y_{l^{\prime}m^{\prime}}(\pi-\Theta,\varphi)\text{ \ \ \ for }%
R<d,\label{ad3b}\\
h_{l}^{(1)}(\omega R)Y_{lm}(\pi-\Theta,\varphi)  &  =\sum\limits_{l^{\prime
}m^{\prime}}\alpha_{l^{\prime}m^{\prime}}^{lm}(-\vec{d})j_{l^{\prime}}(\omega
r)Y_{l^{\prime}m^{\prime}}(\theta,\varphi)\text{ \ \ \ for }r<d, \label{ad3c}%
\end{align}
where%

\begin{align}
\alpha_{l^{\prime}m^{\prime}}^{lm+}(\vec{x})  &  \triangleq\sum
\limits_{\lambda\mu}c(lm|l^{\prime}m^{\prime}|\lambda\mu)j_{\lambda}%
(\omega\left\vert x\right\vert )Y_{\lambda\mu}(\hat{x}),\label{alfa}\\
\alpha_{l^{\prime}m^{\prime}}^{lm}(\vec{x})  &  \triangleq\sum\limits_{\lambda
\mu}c(lm|l^{\prime}m^{\prime}|\lambda\mu)h_{\lambda}^{(1)}(\omega\left\vert
x\right\vert )Y_{\lambda\mu}(\hat{x}).\nonumber
\end{align}
\noindent

The coefficients $c(lm|l^{\prime}m^{\prime}|\lambda\mu)$ in terms of 3-j
symbols are:%

\begin{equation}
c(lm|l^{\prime}m^{\prime}|\lambda\mu)=i^{l+\lambda-1}(-1)^{m}[4\pi
(2l+1)(2l^{\prime}+1)(2\lambda+1)]^{1/2}\left(
\begin{array}
[c]{ccc}%
l & l^{\prime} & \lambda\\
0 & 0 & 0
\end{array}
\right)  \left(
\begin{array}
[c]{ccc}%
l & l^{\prime} & \lambda\\
m & -m^{\prime} & \mu
\end{array}
\right)  \label{c}%
\end{equation}

The expansion (\ref{ad3b}) and (\ref{ad3c}) are valid for $R$\ $<d$\ and
$r$\ $<d$, respectively. and they cover region where the wormhole conditions
are imposed: $R$\ $=a$\ and $r=a,$ ($a<d$).

Using (\ref{ad3a}), (\ref{ad3b}) and (\ref{ad3c}), $\Psi_{0},\Psi_{1}$\ and
$\Psi_{2}$ are expressed as a sum of wave functions at translated origin as:%

\begin{align*}
\sum\limits_{lm}A_{lm}\cdot j_{l}(\omega r)Y_{lm}(\theta,\varphi)  &
=\sum\limits_{lm}\bar{A}_{lm}j_{l}(\omega R)Y_{lm}(\Phi,\varphi),\\
\sum\limits_{lm}B_{lm}\cdot h_{l}^{(1)}(\omega r)Y_{lm}(\theta,\varphi)  &
=\sum\limits_{lm}\bar{B}_{lm}j_{l}(\omega r)Y_{lm}(\Phi,\varphi),\\
\sum\limits_{lm}C_{lm}\cdot h_{l}^{(1)}(\omega R)Y_{lm}(\Phi,\varphi)  &
=\sum\limits_{lm}\bar{C}_{lm}j_{l}(\omega r)Y_{lm}(\theta,\varphi).
\end{align*}
where the analogues of the formulas (\ref{t2a}), (\ref{t2b}) and (\ref{t2c})
are (see appendix A)

\begin{align}
\bar{A}_{lm}  &  =(-1)^{l+m}\sum\limits_{l^{\prime}}A_{l^{\prime}m}\alpha
_{lm}^{l^{\prime}m+}(\vec{d}),\label{3a1}\\
\bar{B}_{lm}  &  =(-1)^{l+m}\sum\limits_{l^{\prime}}B_{l^{\prime}m}\alpha
_{lm}^{l^{\prime}m}(\vec{d}),\label{3b1}\\
\bar{C}_{lm}  &  =(-1)^{l+m}\sum\limits_{l^{\prime}}C_{l^{\prime}m}\alpha
_{lm}^{l^{\prime}m}(\vec{d}). \label{3c1}%
\end{align}

3-j symbols are zero unless $m-m^{\prime}=\mu$ \cite{edmonds}. Furthermore,
$\vec{d}=\hat{z}d,$ and $Y_{\lambda\mu}(\hat{d})=Y_{\lambda\mu}(0,\varphi)$ is
nonzero only when $\mu=0.$ Thus $m^{\prime}=m$ and that's why the summation
over $m^{\prime}$ drops in (\ref{3a1}), (\ref{3b1}) and (\ref{3c1})
\begin{align*}
Y_{\lambda0}(0,\varphi)  &  =\sqrt{\dfrac{2\lambda+1}{4\pi}},\\
\alpha_{lm}^{l^{\prime}m^{\prime}}(\vec{d})  &  =\alpha_{lm}^{l^{\prime}%
m}(\vec{d})=\sum\limits_{\lambda,\mu}c(l^{\prime}m|lm|\lambda0)h_{\lambda
}^{(1)}(\omega d)\sqrt{\dfrac{2\lambda+1}{4\pi}},\\
\alpha_{lm}^{l^{\prime}m^{\prime+}}(\vec{d})  &  =\alpha_{lm}^{l^{\prime}%
m^{+}}(\vec{d})=\sum\limits_{\lambda,\mu}c(l^{\prime}m|lm|\lambda0)j_{\lambda
}(\omega d)\sqrt{\dfrac{2\lambda+1}{4\pi}}.
\end{align*}

Imposing the wormhole conditions and using the orthogonality of $Y_{lm}%
(\theta,\varphi)$ for different $l,m$, yields the 3+1 dimensional analogues of
the equations found for 2+1 dimensions:%

\begin{align*}
B_{lm}-e^{i\omega\tau}C_{lm}  &  =-\dfrac{j_{l}(\omega a)}{h_{l}^{(1)}(\omega
a)}(\bar{C}_{lm}-e^{i\omega\tau}\bar{B}_{lm}+A_{lm}-e^{i\omega\tau}\bar
{A}_{lm}),\\
B_{lm}+e^{i\omega\tau}C_{lm}  &  =-\dfrac{\dfrac{\partial}{\partial r}%
j_{l}(\omega r)|_{r=a}}{\dfrac{\partial}{\partial r}h_{l}^{(1)}(\omega
r)|_{r=a}}(\bar{C}_{lm}-e^{i\omega\tau}\bar{B}_{lm}+A_{lm}-e^{i\omega\tau}%
\bar{A}_{lm}),
\end{align*}
giving%

\begin{align}
B_{lm}  &  =-\gamma_{n}^{+}(\omega a)\bar{C}_{lm}+e^{i\omega\tau}\gamma
_{n}^{-}(\omega a)\bar{B}_{lm}+E_{lm},\label{3da}\\
C_{lm}  &  =-\gamma_{n}^{+}(\omega a)\bar{B}_{lm}+e^{-i\omega\tau}\gamma
_{n}^{-}(\omega a)\bar{C}_{lm}+F_{lm}, \label{3db}%
\end{align}
where $E_{lm}$ and $F_{lm}$ are known functions of $A_{lm}:$%

\begin{align*}
E_{lm}  &  =-\gamma_{n}^{+}(\omega a)A_{lm}+e^{i\omega\tau}\gamma_{n}%
^{-}(\omega a)\bar{A}_{lm},\\
F_{lm}  &  =-\gamma_{n}^{+}(\omega a)\bar{A}_{lm}+e^{-i\omega\tau}\gamma
_{n}^{-}(\omega a)A_{lm},
\end{align*}
and $\gamma_{l}^{\pm}(\omega a)$ are defined similar to 2+1 dimensional case:%

\begin{align*}
\gamma_{l}^{+}(\omega a)  &  \triangleq\dfrac{1}{2}\left(  \dfrac{j_{l}(\omega
a)}{h_{l}^{(1)}(\omega a)}+\dfrac{\dfrac{\partial}{\partial r}j_{l}(\omega
r)|_{r=a}}{\dfrac{\partial}{\partial r}h_{l}^{(1)}(\omega r)|_{r=a}}\right)
,\\
\ \gamma_{l}^{-}(\omega a)  &  \triangleq\dfrac{1}{2}\left(  \dfrac
{j_{l}(\omega a)}{h_{l}^{(1)}(\omega a)}-\dfrac{\dfrac{\partial}{\partial
r}j_{l}(\omega r)|_{r=a}}{\dfrac{\partial}{\partial r}h_{l}^{(1)}(\omega
r)|_{r=a}}\right)  .
\end{align*}

Similar to the 2+1 dimensional case, (\ref{3da}) and (\ref{3db}) can be solved
for $a\ll1$ and $a\ll d$ cases.

\subsection{Solutions for $a\ll d\ $and $a\ll1:$}

The asymptotic form of $h_{l}^{(1)}(\omega d)$ for $l\ll\omega d$ allows us to
compute $\alpha_{lm}^{l^{\prime}m^{\prime}}(\vec{d})$. The similarity between
2+1 and 3+1 dimensional cases are remarkable. Indeed for 2+1 dimensional case,
if we consider $\bar{X}_{n}=\sum\limits_{k=-\infty}^{\infty}X_{k-n}H_{k}%
^{(1)}(\omega d)$ as an operator on $H_{n}^{(1)}(\omega d)$, the asymptotic
form of $H_{n}^{(1)}(\omega d)$ for $n\ll\omega d$ is an eigenvalue of this
operator. Similarly in the passage to the 3+1 dimensions, considering $\bar
{X}_{lm}=(-1)^{l+m}\sum\limits_{l^{\prime}m^{\prime}}X_{l^{\prime}m^{\prime}%
}\alpha_{lm}^{l^{\prime}m^{\prime}}(\vec{d})$ as an operator on $h_{l}%
^{(1)}(\omega d)$, asymptotic form of $h_{l}^{(1)}(\omega d)$ for $l\ll\omega
d$ is an eigenfunction of $\bar{X}_{lm}$.

As in the 2+1 dimensional case, the presence of the $\gamma_{n}^{\pm}(\omega
a)$ factor at each term of the right hand sides of (\ref{3da}) and
(\ref{3db}), makes $B_{lm}$ and $C_{lm}$ vanish when $\omega a\ll l.$ Thus
when $a\ll d$ the asymptotic form of $h_{\lambda}^{(1)}(\omega d)$ for
$l\ll\omega d$ can be used.

For $a\ll1,$ just like 2+1 dimensions, $h_{l}^{(1)}(\omega d)$ is zero unless
$l=0$ and (\ref{3da}) and (\ref{3db}) can be solved.

\subsubsection{$a\ll d$:}

$\gamma_{l}^{+}(\omega a)$ and $\gamma_{l}^{-}(\omega a)\ $filter the terms
with $l>2\omega a,$ thus when $a\ll d$ the only terms that contribute to
$\bar{B}_{lm}$ and $\bar{C}_{lm}$are $l\ll\omega d.$ In this case $h_{\lambda
}^{(1)}(\omega d)$ has the asymptotic expression:%

\[
h_{\lambda}^{(1)}(\omega d)\approx i^{-(\lambda+1)}\dfrac{e^{i\omega d}%
}{\omega d}%
\]

Then,%

\[
\bar{B}_{lm}\approx\sum\limits_{l^{\prime}m^{\prime}}\sum\limits_{\lambda
}B_{l^{\prime}m^{\prime}}c(l^{\prime}m|lm|\lambda0)i^{-(\lambda+1)}%
\dfrac{e^{i\omega d}}{\omega d}\sqrt{\dfrac{2\lambda+1}{4\pi}}%
\]

Substituting%

\[
c(l^{\prime}m|lm|\lambda0)=i^{l^{\prime}+\lambda-1}(-1)^{m}[4\pi
(2l+1)(2l^{\prime}+1)(2\lambda+1)]^{1/2}\left(
\begin{array}
[c]{ccc}%
l & l^{\prime} & \lambda\\
0 & 0 & 0
\end{array}
\right)  \left(
\begin{array}
[c]{ccc}%
l & l^{\prime} & \lambda\\
m & -m & 0
\end{array}
\right)
\]
gives:%
\[
\bar{B}_{lm}\approx-\dfrac{e^{i\omega d}}{\omega d}\sum\limits_{l^{\prime}%
}B_{l^{\prime}m}i^{l^{\prime}}(-1)^{m}[(2l+1)(2l^{\prime}+1)]^{1/2}\delta_{m0}%
\]
where in the last step the orthogonality property of the 3-j symbols is used
\cite{talman}:%

\[
\sum\limits_{\lambda\mu}(2\lambda+1)\left(
\begin{array}
[c]{ccc}%
l & l^{\prime} & \lambda\\
m_{1} & m_{2} & \mu
\end{array}
\right)  \left(
\begin{array}
[c]{ccc}%
l & l^{\prime} & \lambda\\
p_{1} & p_{2} & \mu
\end{array}
\right)  =\delta_{m1p_{1}}\delta_{m2p2}.
\]

Thus,%

\[
\bar{B}_{lm}\approx-\dfrac{e^{i\omega d}}{\omega d}\sqrt{(2l+1)}%
\sum\limits_{l^{\prime}}B_{l^{\prime}0}i^{l^{\prime}}\sqrt{(2l^{\prime}%
+1)}\delta_{m0}=-\dfrac{e^{i\omega d}}{\omega d}\sqrt{(2l+1)}T(B_{l0}%
)\delta_{m0},
\]
where, for $X_{l}$ being any function of $l$, the functional $T(X_{l})$ is
defined as:%

\[
T(X_{l})\triangleq\sum\limits_{l^{\prime}}X_{l^{\prime}}i^{l^{\prime}}%
\sqrt{(2l^{\prime}+1)}%
\]

If $m\neq0;$ $B_{lm}=E_{lm},C_{lm}=F_{lm}$ and if $m=0:$%

\begin{align*}
B_{l0}  &  =-e^{i\omega\tau}\dfrac{e^{i\omega d}}{\omega d}(-1)^{l}%
\sqrt{(2l+1)}\gamma_{l}^{-}(\omega a)T(B_{l0})+\dfrac{e^{i\omega d}}{\omega
d}(-1)^{l}\sqrt{(2l+1)}\gamma_{l}^{+}T(C_{l0})+E_{l0},\\
C_{l0}  &  =-e^{-i\omega\tau}\dfrac{e^{i\omega d}}{\omega d}(-1)^{l}%
\sqrt{(2l+1)}\gamma_{l}^{-}(\omega a)T(C_{l0})+\dfrac{e^{i\omega d}}{\omega
d}(-1)^{l}\sqrt{(2l+1)}\gamma_{l}^{+}T(B_{l0})+F_{l0}.
\end{align*}

Multiplying each side of these equations by $i^{l}\sqrt{(2l+1)}$ and summing
over $l$ gives $T(B_{l0})$ and $T(C_{l0}):$%

\[%
\begin{bmatrix}
T(B_{l0})\\
T(C_{l0})
\end{bmatrix}
=%
\begin{bmatrix}
1-e^{i\omega\tau}\dfrac{e^{i\omega d}}{\omega d}T((-i)^{l}(2l+1)\gamma_{l}%
^{-}(\omega a)) & \dfrac{e^{i\omega d}}{\omega d}T((-i)^{l}(2l+1)\gamma
_{l}^{+}(\omega a))\\
\dfrac{e^{i\omega d}}{\omega d}T((-i)^{l}(2l+1)\gamma_{l}^{+}(\omega a)) &
1-e^{-i\omega\tau}\dfrac{e^{i\omega d}}{\omega d}T((-i)^{l}(2l+1)\gamma
_{l}^{-}(\omega a))
\end{bmatrix}
^{-1}%
\begin{bmatrix}
T(E_{l0})\\
T(F_{l0})
\end{bmatrix}
\]

\subsubsection{$a\ll1$:}

Similar to the 2+1 dimensional case, for $a\ll1$, $\gamma_{l}^{\pm}(\omega a)$
becomes a discrete delta function, $\delta_{l}.$ Due to the factors of
$\gamma_{l}^{\pm}(\omega a)$ in each term$,$ $B_{lm}$ and $C_{lm}$ are nonzero
for only $l=m=0.$ The problem reduces to finding the constants $B_{00}$ and
$C_{00}.$%

\[
\gamma_{l}^{+}(\omega a)\approx\gamma_{l}^{-}(\omega a)\approx\dfrac{\omega
a}{i+\omega a}\delta_{l},
\]

\begin{align*}
B_{lm}  &  =B_{00}\delta_{l}\delta_{m},\\
C_{lm}  &  =C_{00}\delta_{l}\delta_{m}.
\end{align*}

$l=0$ implies $m=0$ and $l^{\prime}=\lambda,$ so that%

\begin{align*}
\bar{B}_{00}  &  =\sum\limits_{\lambda}B_{00}\delta_{\lambda}c(\lambda
0|00|\lambda0)h_{\lambda}^{(1)}(\omega d)\sqrt{\dfrac{2\lambda+1}{4\pi}%
}=B_{00}h_{0}^{(1)}(\omega d),\\
\bar{C}_{00}  &  =\sum\limits_{\lambda}C_{00}\delta_{\lambda}c(\lambda
0|00|\lambda0)h_{\lambda}^{(1)}(\omega d)(-1)^{\lambda}\sqrt{\dfrac
{2\lambda+1}{4\pi}}=C_{00}h_{0}^{(1)}(\omega d).
\end{align*}

$B_{00}$ and $C_{00}$ are found as:%

\[%
\begin{bmatrix}
B_{00}\\
C_{00}%
\end{bmatrix}
=%
\begin{bmatrix}
1-e^{i\omega\tau}h_{0}^{(1)}(\omega d) & h_{0}^{(1)}(\omega d)\\
h_{0}^{(1)}(\omega d) & 1-e^{-i\omega\tau}h_{0}^{(1)}(\omega d)
\end{bmatrix}
^{-1}%
\begin{bmatrix}
E_{00}\\
F_{00}%
\end{bmatrix}
.
\]

\subsection{Multiple scattering}

Multiple scattering formulae for the 3+1 dimensions can be found. by the same
steps followed as the 2+1 dimensional case The multiple scattering expansion
of 3+1 dimensional wave functions%

\begin{align*}
\Psi_{1}^{k}  &  =\sum\limits_{lm}B_{lm}^{k}\cdot h_{l}^{(1)}(\omega
r)Y_{lm}(\theta,\varphi),\\
\Psi_{2}^{k}  &  =\sum\limits_{lm}C_{lm}^{k}\cdot h_{l}^{(1)}(\omega
R)Y_{lm}(\Theta,\varphi),
\end{align*}
together with the wormhole conditions for the $1^{st}$ and the $k^{th}$ order
scattering coefficients%

\begin{align*}
(\Psi_{0}+\Psi_{1}^{1})_{r=a}  &  =e^{i\omega\tau}(\Psi_{0}+\Psi_{2}%
^{1})_{R=a,\Theta=\theta},\\
\dfrac{\partial}{\partial r}(\Psi_{0}+\Psi_{1}^{1})|_{r=a}  &  =-e^{i\omega
\tau}\dfrac{\partial}{\partial R}(\Psi_{0}+\Psi_{2}^{1})|_{R=a,\Theta=\theta
},\\
(\Psi_{1}^{k+1}+\Psi_{2}^{k})|_{r=a}  &  =e^{i\omega\tau}(\Psi_{1}^{k}%
+\Psi_{2}^{k+1})|_{R=a,\Theta=\theta},\\
\dfrac{\partial}{\partial r}(\Psi_{1}^{k+1}+\Psi_{2}^{k})|_{r=a}  &
=-e^{i\omega\tau}\dfrac{\partial}{\partial R}(\Psi_{1}^{k}+\Psi_{2}%
^{k+1})|_{R=a,\Theta=\theta},
\end{align*}
lead the formulas for the multiple scattering solution:%

\begin{align*}
B_{lm}^{1}  &  =-\gamma_{l}^{+}(\omega a)A_{lm}+e^{i\omega\tau}\gamma_{l}%
^{-}(\omega a)\bar{A}_{lm},\\
C_{lm}^{1}  &  =-\gamma_{l}^{+}(\omega a)\bar{A}_{lm}+e^{-i\omega\tau}%
\gamma_{n}^{-}(\omega a)A_{lm},\\
B_{lm}^{k+1}  &  =-\gamma_{l}^{+}(\omega a)\bar{C}_{lm}^{k}+e^{i\omega\tau
}\gamma_{n}^{-}(\omega a)\bar{B}_{lm}^{k},\\
C_{lm}^{k+1}  &  =-\gamma_{l}^{+}(\omega a)\bar{B}_{lm}^{k}+e^{-i\omega\tau
}\gamma_{n}^{-}(\omega a)\bar{C}_{lm}^{k}.
\end{align*}

\section{Numerical verifications:}

In this section, the solutions for certain values of $a,\ d,\ \omega\ $and
$\tau$ are evaluated numerically for 2+1 dimensions and\ it is verified that
they satisfy wormhole conditions. Numerical evaluation of solutions are done
by using the multiple scattering results (\ref{ms2a1}), (\ref{ms2b1}),
(\ref{ms2ak}) and (\ref{ms2bk}). Alternatively (\ref{2da}) and (\ref{2db}) are
tested by an iteration method. For iteration, two initial test functions
$B_{n}^{0}$ and $C_{n}^{0}$ are picked and substituted to right hand sides of
(\ref{2da}) and (\ref{2db}) to obtain $B_{n}^{1}$ and $C_{n}^{1}$. Similarly
$B_{n}^{1}$ and $C_{n}^{1}$ are substituted to (\ref{2da}) and (\ref{2db}) to
obtain $B_{n}^{2}$ and $C_{n}^{2}.$ Continuing this iteration, $B_{n}^{m}$ and
$C_{n}^{m}$ are assumed to converge to the solution. No proof for the
conditions of convergence is given, it is verified numerically that the
solution found by iteration method converges to the multiple scattering
solution for the parameter sets that are considered.%

\begin{figure}
[ptb]
\begin{center}
\includegraphics[
width=3.7273in
]%
{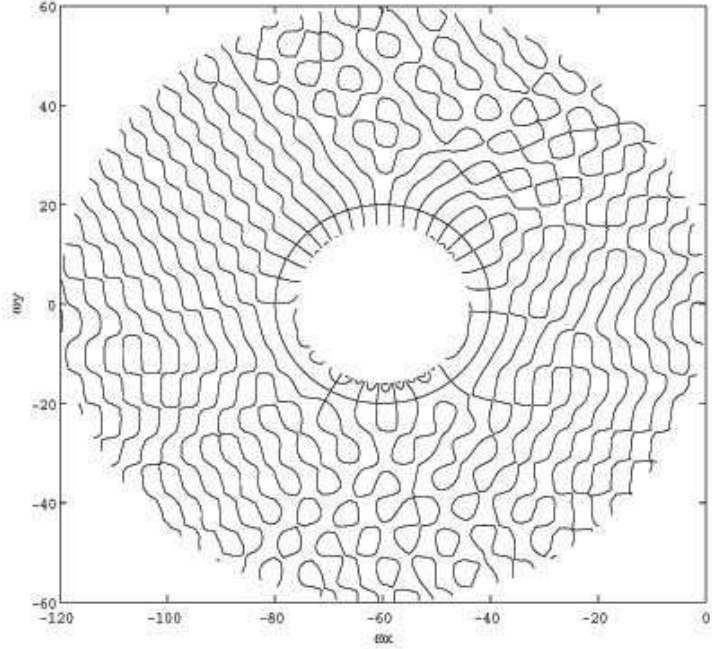}%
\caption{The contour plot of $\operatorname{Re}(\Psi-e^{i\omega\tau}R\Psi)$ in
the vicinity of left wormhole mouth $\Delta_{-}.$ The contour circle at $r=a$
shows that $(\Psi-e^{i\omega\tau}R\Psi)|_{r=a}$ is constant. $(\omega
a=20;\ \omega d=120;\ \tau=1;\ \alpha=\pi/3)$}%
\end{center}
\end{figure}

\bigskip%

\begin{figure}
[ptb]
\begin{center}
\includegraphics[
width=3.6858in
]%
{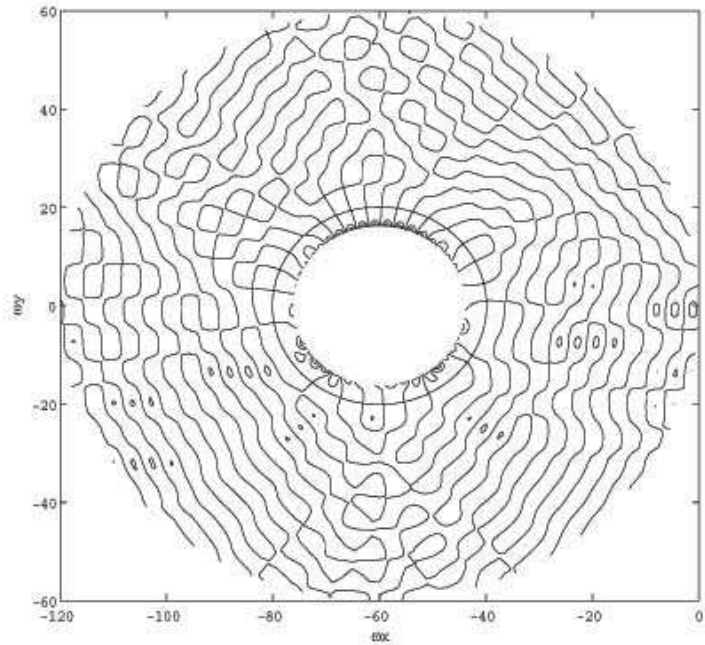}%
\caption{The contour plot of $\operatorname{Re}(\dfrac{\partial}{\partial
r}(\Psi+e^{i\omega\tau}R\Psi))$ in the vicinity of left wormhole mouth
$\Delta_{-}.$ The same contour circle at $r=a$ is evident.$(\omega
a=20;\ \omega d=120;\ \tau=1;\ \alpha=\pi/3)$}%
\end{center}
\end{figure}

Moreover, to check the formulas found for $a\ll d$ the solutions found by this
method is compared with the multiple scattering solution.

As the velocity of wave is taken as $1$ in equation (\ref{helm}), $\omega
d=2\pi d/\lambda$ and $\omega a=2\pi a/\lambda$ where $\lambda$ is the
wavelength of the wave. Practically if when a light wave in a wormhole
universe is considered, these values supposed to be much larger (at least
order of $\sim10^{10})$ compared to what chosen in the below examples.
However, numerical calculations with such large values were beyond the
capacity of the PC used and there is no reason to think that the formulas will
fail for large values.%

\begin{figure}
[ptb]
\begin{center}
\includegraphics[
width=3.8527in
]%
{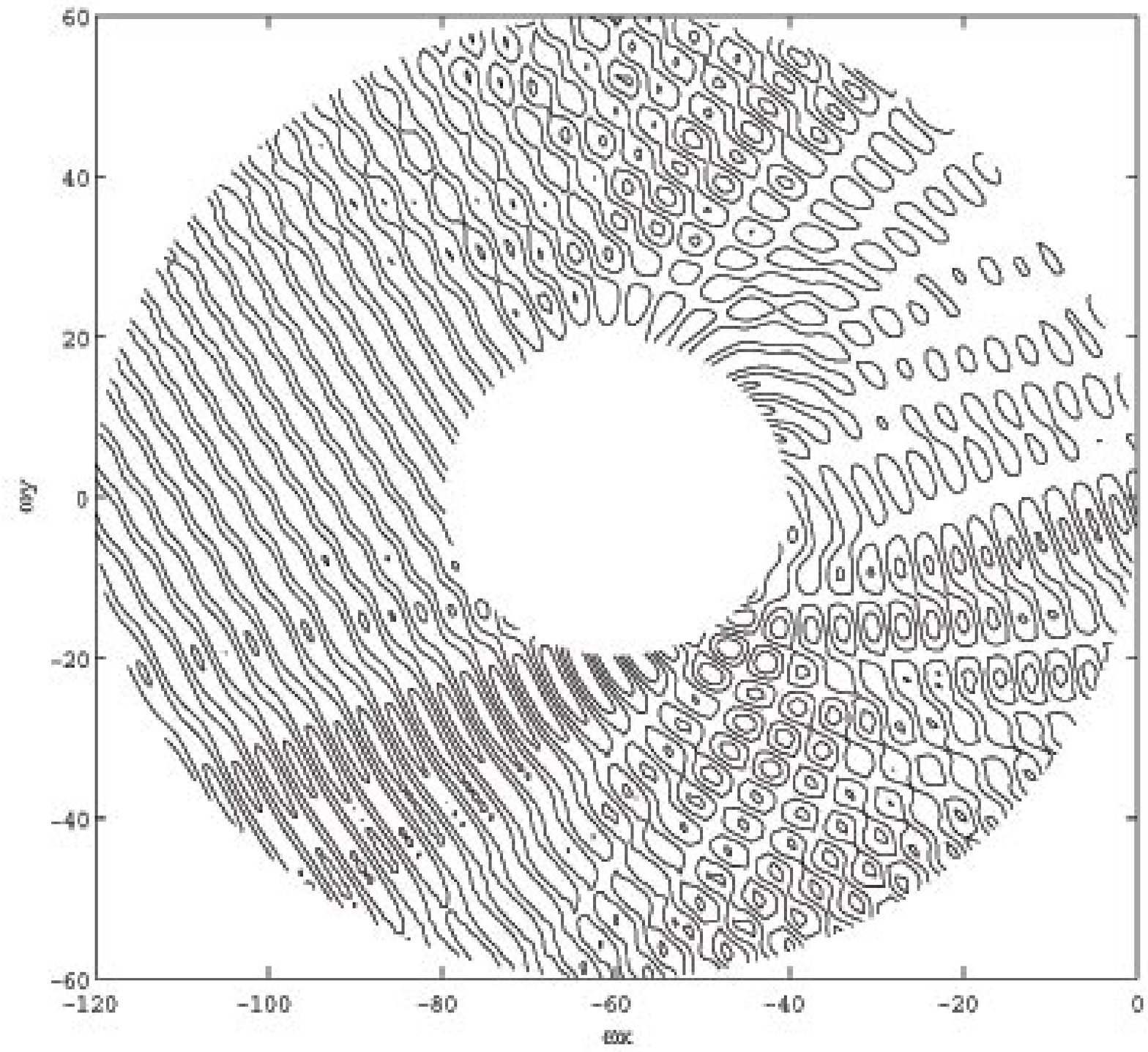}%
\caption{The contour plot of $\operatorname{Re}(\Psi)$ in the vicinity of
$\Delta_{-}.$ The incident wave is coming from the left with an angle $\pi/3$
and the shadow is on the opposite side. $(\omega a=20;\ \omega d=120;\ \tau
=1;\ \alpha=\pi/3).$}%
\end{center}
\end{figure}

The incident wave $\Psi_{0}$ is chosen as a plane wave and $A_{n}=e^{in\alpha
},$ where $\alpha$ is the angle between direction of the incident wave and the
$y$ axis.

Referring to figure $2$, the wormhole is located symmetrically with respect to
the $y$ axis. Consider the reflection operator $R$ with respect to the $y$
axis, i.e. $R\Psi(x,y)=\Psi(-x,y)$. According to the wormhole conditions C-1
and C-2%

\begin{align}
(\Psi-e^{i\omega\tau}R\Psi)|_{r=a}  &  =0\label{test1}\\
\dfrac{\partial}{\partial r}(\Psi+e^{i\omega\tau}R\Psi)|_{r=a}  &  =0
\label{test2}%
\end{align}

It is verified that the solution found satisfies (\ref{test1})and
(\ref{test2}) by plotting the contours at the vicinity of one of the wormhole mouths.%

\begin{figure}
[ptb]
\begin{center}
\includegraphics[
width=4.1667in
]%
{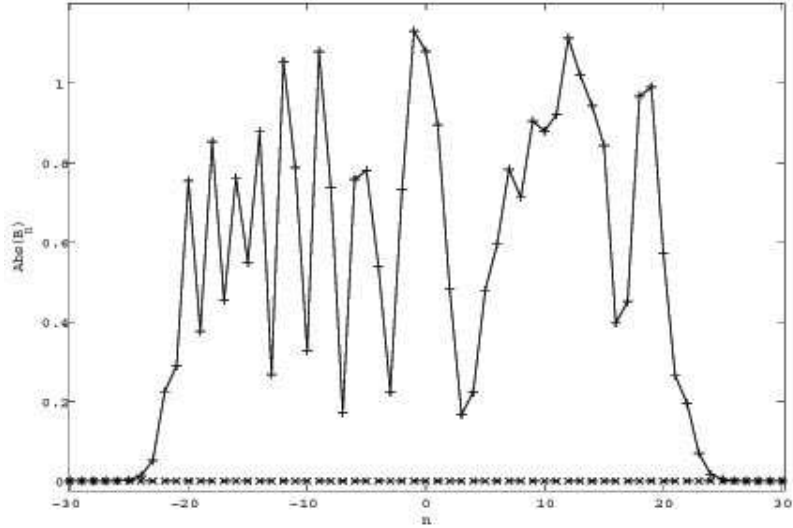}%
\caption{Comparison of the multiple scattering and the iteration results. The
difference of $|B_{n}|$ found by these two methods are points with marker `x'
which are zero for all $n.$ $(\omega a=20;\ \omega d=120;\ \tau=1;\ \alpha
=\pi/3)$}%
\end{center}
\end{figure}

In fig.4, fig. 5, fig. 6 and fig. 7 the parameters are: $\omega a=20,\ \omega
d=120,\ \alpha=\pi/3,\ \omega\tau=1.$Figure 4 and Figure 5 show contour plots
of the multiple scattering solution for real part of $\Psi(x,y)+e^{i\omega
\tau}R(\Psi(x,y)$ and $\dfrac{\partial}{\partial r}(\Psi(x,y)-e^{i\omega\tau
}R(\Psi(x,y)),$ respectively$.$ In both figures, the contour circles at
$\omega r=\omega a=20$ are clearly visible indicating that the values of each
function are zero along $r=a$ circle. This shows that the wormhole conditions
are satisfied. The contour plots of imaginary parts -which are not presented
here- give the same contour circles at $\ r=a.$ Although the contour is
plotted for $0.8a<r<d$ to make the zero contour circle more visible, it should
be remembered that the region $r<a$ is not a part of the spacetime. Figure 6
is a contour plot of the real part of the solution $\Psi(r,\theta)$ to give an
example of a visual image of the solution. Figure 7 is a comparison of the
multiple scattering solution and the iteration solution. The solid line with
`+' markers show the $|B_{n}|$ that are found by multiple scattering and
dashed line with `x' markers are the difference of the\ absolute values of
$B_{n}$ found by the iteration method and the multiple scattering method. The
difference is zero for all $n$; i.e. these two solutions are exactly the same.
The results are obtained after 20 iterations. The test functions are chosen as
constant, $B_{n}^{0}=C_{n}^{0}=1.$

In fig. 8, the parameters are: $\omega a=5,\ \omega d=1600,\ \alpha
=\pi/5,\ \omega\tau=1.$ This is an example for $a\ll d$ case. In the figure
$|B_{n}|$ versus $n$ is plotted. The solid line with `+' markers is the
multiple scattering solution and the dashed line with `x' markers is the $a\ll
d$ approximation solutions given by (\ref{agd_b}) and (\ref{agd_c}).%

\begin{figure}
[ptb]
\begin{center}
\includegraphics[
width=4.222in
]%
{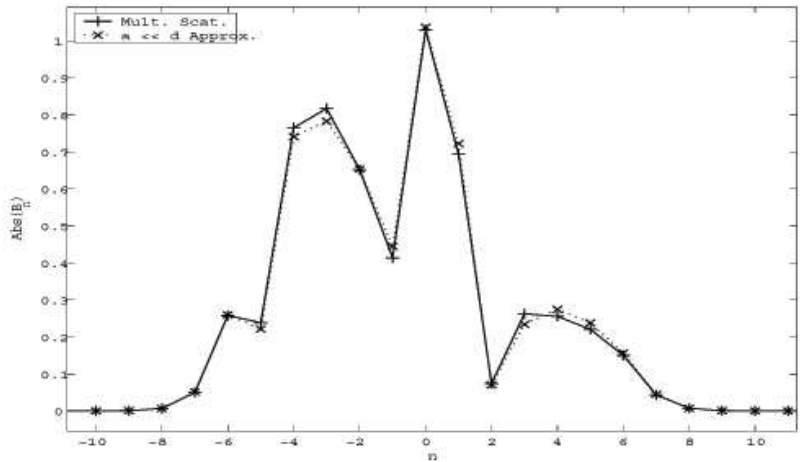}%
\caption{Comparison of the multiple scattering and the $a\ll d$ approximation.
$(\omega a=5;\ \omega d=1600;\ \tau=1;\ \alpha=\pi/5)$}%
\end{center}
\end{figure}

\section{Concluding Remarks}

The principal purpose of the present work was to investigate the scalar waves
in wormhole topology. As the wormhole considered is flat, no complications due
to curvature arise.

Although the spacetime considered in this work admits CTCs for sufficiently
large values of time lag $\tau$, their existence has no influence on
monochromatic waves. The closed timelike curves emerge when time lag $\tau$ is
greater than $d-2a$. However, $\tau$ appears in the equations only as
$\exp(i\omega\tau).$ Thus the solution remains the same for all integer $k$'s
where $\omega\tau=2k\pi+\alpha$ and increasing $\tau$ does not change the
nature of the solutions. This suggests that the presence of closed timelike
curves does not have a dramatic effect on the scalar wave solutions.

This should not be surprising considering that, in a wave equation, what
really matters is presence of closed null curves, rather than closed timelike
curves. It is reasonable to think that the existence of closed timelike curves
will not effect the nature of the solutions as long as closed null curves are
not present. CTC's are present in the flat wormhole spacetime studied here,
but still they don't have a significant effect on the solution. The reason is
explained in \cite{FM}: In this kind of spacetimes, the closed null curves are
a set of measure zero and due to the diverging lens property of the wormhole,
the strength of the field is weakened by a factor $a/2d$ at each loop in the
infinitely looping closed null geodesics.

The complications related to closed timelike curves are due to difficulty in
specifying a Cauchy hypersurface when solving the Cauchy problem. Null
geodesics are bicharacteristics of the wave equation and arbitrary initial
data cannot be properly posed in a null direction \ \cite{fried}. A spacelike
hypersurface never contains vectors in a null direction, thus are good
candidates for specifying initial data. However there always exist a null
direction on a timelike point of a hypersurface. In the light of these
discussions it can be conjectured that no complications arise on the solution
of wave equation due to CTC's. The complications are mainly due to the nature
of Cauchy problem approach.

To make this point more apparent, let's consider a space-time with topology
$\mathbb{R}^{2}\times S^{1},$ where $S^{1}$ is the temporal direction. Clearly
this spacetime admits closed timelike curves. In this spacetime, waves from
only a discrete set of frequencies can exist due to periodicity conditions.
But this restriction does not seem to be related to the existence of CTC's,
because alternatively when a globally hyperbolic spacetime of topology
$S^{2}\times\mathbb{R}$ is considered, it is easy to see that it has the same
property: Waves only from a discrete set of frequencies can exist in this
spacetime, either. This observation also suggests that the restrictions on the
solutions of wave equation are not due to existence of closed timelike curves.
Compactness of the topology in either temporal or spatial directions have both
similar consequences. Thus it is reasonable to expect that waves in a universe
where time travel is possible, will have similar properties with globally
hyperbolic ones.

If we consider the question in a purely mathematical point of view, the form
of wave equation considered is almost symmetric with respect to time and space
variables. For the 1+1 dimensions there is complete symmetry (remembering that
the minus sign on the time derivative does not effect the symmetry since it is
always possible to reverse the signs) and for higher dimensions the only
difference is having more space variables. This suggest that there is no
strong mathematical background for expecting disparate consequences of
existence of CTC's compared to existence of closed curves along any space
direction. On the other hand more space coordinates give rise to asymmetry
between the spacelike hypersurfaces and the timelike hypersurfaces in Cauchy
problem due to the shape of the null cone: Any timelike hypersurface passing
through a spacetime point intersects the null cone of that point, while
spacelike hypersurfaces does not.

There is a strong analogy between 2+1 and 3+1 cases, which suggests that the
results can be extended to $n+1$ dimensions easily. In any dimensions, the
solutions can be expressed in terms of spherical waves, $f(r)Y(\Omega)$, where
$r$ is the radial distance and $\Omega$ denotes the angular part
\cite{watson}. In addition, to be able to apply the same method, an addition
theorem similar to that of the 2+1 and 3+1 dimensions is needed for this
higher dimension. The similarity of (\ref{2da}), (\ref{2db}) with (\ref{3da}),
(\ref{3db}) suggests that the solution for higher dimensions are readily given
by these equations where the expressions of $\bar{B}$ and $\bar{C}$ in terms
of $B$ and $C$ will be found using addition theorems of those dimensions.

\begin{acknowledgement}
The author thanks Cem Tezer for his valuable guidance\ and helpful
discussions, John L. Friedman for his significant comments, and Ozgur Sarioglu.
\end{acknowledgement}

\bigskip\appendix

\section{Calculation of $\bar{A}_{lm,}\ \bar{B}_{lm}$ and $\bar{C}_{lm}$}

Referring to the equations (\ref{3jadd}), (\ref{3hr}) and (\ref{3hR})%

\begin{align*}
\sum\limits_{lm}A_{lm}\cdot j_{l}(\omega r)Y_{lm}(\theta,\varphi)  &
=\sum\limits_{lm}\bar{A}_{lm}j_{l}(\omega R)Y_{lm}(\Phi,\varphi),\\
\sum\limits_{lm}B_{lm}\cdot h_{l}^{(1)}(\omega r)Y_{lm}(\theta,\varphi)  &
=\sum\limits_{lm}\bar{B}_{lm}j_{l}(\omega r)Y_{lm}(\Phi,\varphi),\\
\sum\limits_{lm}C_{lm}\cdot h_{l}^{(1)}(\omega R)Y_{lm}(\Phi,\varphi)  &
=\sum\limits_{lm}\bar{C}_{lm}j_{l}(\omega r)Y_{lm}(\theta,\varphi),
\end{align*}
(\ref{ad3a}), (\ref{ad3b}) and (\ref{ad3c}) can be employed to calculate
$\bar{A}_{lm}$, $\bar{B}_{lm}$ and $\bar{C}_{lm}.$ Considering $\Psi_{0},$%

\begin{align*}
\Psi_{0}  &  =\sum\limits_{lm}A_{lm}\cdot j_{l}(\omega r)Y_{lm}(\theta
,\varphi)=\sum\limits_{lm}A_{lm}\sum\limits_{l^{\prime}m^{\prime}}%
\alpha_{l^{\prime}m^{\prime}}^{lm+}(\vec{d})j_{l^{\prime}}(\omega
R)Y_{l^{\prime}m^{\prime}}(\pi-\Theta,\varphi)\\
&  =\sum\limits_{lm}A_{lm}\sum\limits_{l^{\prime}m^{\prime}}\alpha_{l^{\prime
}m^{\prime}}^{lm+}(\vec{d})j_{l^{\prime}}(\omega R)(-1)^{l^{\prime}+m^{\prime
}}Y_{l^{\prime}m^{\prime}}(\Theta,\varphi)\\
&  =\sum\limits_{lm}(-1)^{l+m}(\sum\limits_{l^{\prime}m^{\prime}}A_{l^{\prime
}m^{\prime}}\alpha_{lm}^{l^{\prime}m^{\prime}+}(\vec{d}))j_{l}(\omega
R)Y_{lm}(\Theta,\varphi).
\end{align*}

In the last step the order of summations and indices $lm$ and $l^{\prime
}m^{\prime}$ are interchanged. It is assumed that these series converge and
changing the order of the summations is valid \cite{russek}.

Thus%

\[
\bar{A}=(-1)^{l+m}(\sum\limits_{l^{\prime}m^{\prime}}A_{l^{\prime}m^{\prime}%
}\alpha_{lm}^{l^{\prime}m^{\prime}+}(\vec{d})).
\]

Calculation of the $\bar{B}_{lm}$ is identical except $\alpha_{lm}^{l^{\prime
}m^{\prime}+}$ is replaced by $\alpha_{lm}^{l^{\prime}m^{\prime}}.$ To find
$C_{lm}$:

\begin{align}
\Psi_{2}  &  =\sum\limits_{lm}C_{lm}\cdot h_{l}^{(1)}(\omega R)Y_{lm}%
(\Theta,\varphi)=\sum\limits_{lm}(-1)^{l+m}C_{lm}\cdot h_{l}^{(1)}(\omega
R)Y_{lm}(\pi-\Theta,\varphi)\nonumber\\
&  =\sum\limits_{lm}(-1)^{l+m}C_{lm}\sum\limits_{l^{\prime}m^{\prime}}%
\alpha_{l^{\prime}m^{\prime}}^{lm}(-\vec{d})j_{l^{\prime}}(\omega
r)Y_{l^{\prime}m^{\prime}}(\theta,\varphi)\nonumber\\
&  =\sum\limits_{lm}(\sum\limits_{l^{\prime}m^{\prime}}(-1)^{l^{\prime
}+m^{\prime}}C_{l^{\prime}m^{\prime}}\alpha_{lm}^{l^{\prime}m^{\prime}}%
(-\vec{d}))j_{l}(\omega r)Y_{lm}(\theta,\varphi)\nonumber\\
&  =\sum\limits_{lm}(\sum\limits_{l^{\prime}m^{\prime}}(-1)^{l^{\prime
}+m^{\prime}}C_{l^{\prime}m^{\prime}}\alpha_{lm}^{l^{\prime}m^{\prime}}%
(-\vec{d}))j_{l}(\omega r)Y_{lm}(\theta,\varphi). \label{apy}%
\end{align}

Recalling (\ref{alfa}):%

\begin{align*}
\alpha_{lm}^{l^{\prime}m^{\prime}}(-\vec{d})  &  =\alpha_{lm}^{l^{\prime}%
m}(-\vec{d})\sum\limits_{\lambda\mu}c(lm|l^{\prime}m|\lambda0)h_{\lambda
}^{(1)}(\omega d)Y_{\lambda0}(\pi,\varphi)\\
&  =\sum\limits_{\lambda\mu}c(lm|l^{\prime}m|\lambda0)h_{\lambda}^{(1)}(\omega
d)(-1)^{\lambda}\sqrt{\dfrac{2\lambda+1}{4\pi}}.
\end{align*}

The $\left(
\begin{array}
[c]{ccc}%
l & l^{\prime} & \lambda\\
0 & 0 & 0
\end{array}
\right)  $ factor in $c(lm|l^{\prime}m|\lambda0)$ is zero when $l^{\prime
}+l+\lambda$ is odd. Thus $(-1)^{l+l^{\prime}+\lambda}=1$ and $(-1)^{\lambda
}=(-1)^{l+l^{\prime}},$ yielding%

\[
\alpha_{lm}^{l^{\prime}m^{\prime}}(-\vec{d})=(-1)^{l+l^{\prime}}%
\sum\limits_{\lambda\mu}c(lm|l^{\prime}m|\lambda0)h_{\lambda}^{(1)}(\omega
d)\sqrt{\dfrac{2\lambda+1}{4\pi}}=(-1)^{l+l^{\prime}}\alpha_{lm}^{l^{\prime}%
m}(\vec{d}).
\]

Substituting in (\ref{apy})%

\[
\sum\limits_{lm}C_{lm}\cdot h_{l}^{(1)}(\omega R)Y_{lm}(\Theta,\varphi
)=\sum\limits_{lm}((-1)^{l+m}\sum\limits_{l^{\prime}}C_{l^{\prime}m^{\prime}%
}\alpha_{lm}^{l^{\prime}m}(\vec{d}))j_{l}(\omega r)Y_{lm}(\theta,\varphi).
\]

Hence%

\[
\bar{C}_{lm}=(-1)^{l+m}\sum\limits_{l^{\prime}}C_{l^{\prime}m}\alpha
_{lm}^{l^{\prime}m}(\vec{d}).
\]

\end{document}